\documentclass[pra,aps,groupedaddress,showpacs,twocolumn,floatfix,nofootinbib]{revtex4}
\usepackage{amsmath,amsthm,amsfonts,amssymb,graphics,graphicx,epsfig,color,natbib,subfigure}
\usepackage{array}
\usepackage{bbm}
\usepackage{bm}
\usepackage{combelow}
\usepackage{soul}
\usepackage{ulem}
\newcommand{\xfrac}[2]{{#1}/{#2}}

\begin{document}
\title{Precision bounds for noisy nonlinear quantum metrology}
\author{Marcin Zwierz${}^{1, 2}$}\email{zwierz.marcin@gmail.com}
\author{Howard M. Wiseman${}^{1}$}
\affiliation{${}^1$Centre for Quantum Computation and Communication Technology (Australian Research Council), Centre for Quantum Dynamics, Griffith University, Brisbane, QLD 4111, Australia\\
${}^2$Faculty of Physics, University of Warsaw, 00-681 Warsaw, Poland}
\date{\today}

\begin{abstract}
We derive the ultimate bounds on the performance of nonlinear measurement schemes in the presence of noise.  In particular, we investigate the precision of the second-order estimation scheme in the presence of  the two most detrimental types of noise, photon loss and phase diffusion. We find that the second-order estimation scheme is affected by both types of noise in an analogous way as the linear one. Moreover, we observe that for both types of noise the gain in the phase sensitivity with respect to the linear estimation scheme is given by a multiplicative term $\mathcal{O}(1/N)$. Interestingly, we also find that under certain circumstances, a careful engineering of the environment can, in principle, improve the performance of measurement schemes affected by phase diffusion.
\end{abstract}

\pacs{03.65.Ta, 03.67.--a, 42.50.Lc, 06.20.Dk}
\maketitle

\section{Introduction}\noindent
Quantum metrology is an important branch of science that promises many advances in precision measurements \cite{caves81, giovannetti11}. The importance of quantum metrology stems from the fact that an optimally designed quantum measurement always outperforms the analogous classical measurement. A typical measurement scheme involves a probe system $S$ prepared in an initial quantum state $\rho_{S}(0)$ undergoing a unitary transformation to the state $\rho_{S}(\phi) = e^{-i \phi G} \rho_{S}(0) e^{i \phi G}$, where $G$ is some Hermitian operator generating translations in the parameter $\phi$ we wish to estimate. Most of the measurement schemes employ linear operators to generate parameter shifts. For example, in quantum optics we usually have $G = \hat{n}$, where $\hat{n}$ is the photon number operator. However, more generally, the parameter shift $\phi$ can be generated by an optical nonlinearity \cite{boixo07,napolitano11}. In such a case, the generator $G$ is a nonlinear function of the photon number operator $\hat{n}$ such as $G = \hat{n}^{q}$, where $q$ denotes the order of the nonlinearity. Following the evolution, the probe system $S$ is subjected to a generalized measurement $M$, described by some Positive Operator-Valued Measure (POVM) that consists of elements $M_{x}$,  where $x$ denotes the measurement outcome used to make an estimate $\hat{\phi}$ of value of $\phi$. Given the probability distribution of the measurement data $p(x|\phi) = \text{Tr}[M_{x} \rho_{S}(\phi)]$, the performance of a parameter estimation scheme can be bounded by the quantum Cram\'er-Rao bound
\begin{equation}\label{CRbound}
\Delta \phi \geq \frac{1}{\sqrt{m F_{Q}[\rho_{S}(\phi)]}}\, ,
\end{equation}
where $\Delta \phi = \langle (\hat{\phi} - \phi)^2 \rangle^{1/2}$ is the square root of the mean-square error with the average taken over all possible measurement outcomes, $m$ denotes the number of measurement repetitions and $F_{Q}[\rho_{S}(\phi)]$ is the quantum Fisher information \cite{braunstein94, braunstein96}. We emphasize that the quantum Cram\'er-Rao bound is valid \textit{only} for unbiased estimators, and therefore should be used with caution (for more details on the limitations of the Cram\'er-Rao bound see Sec.~VIII in Ref.~\cite{berry12} and for alternative bounds see Refs.~\cite{hall12,hallnjp12,hallprx12,nair12arXiv}).

It is well known that in the case of linear parameter estimation schemes, the error $\Delta \phi$ scales at best with the Heisenberg-limited sensitivity as $1/N$, where $N$ denotes the average number of photons used in the scheme. This scaling represents a significant $\sqrt{N}$ improvement with respect to the classical shot-noise limit. However, there is one complication: this improvement is only present in \textit{noiseless} measurement schemes. Since all precision measurements suffer from some form of noise it is of the utmost importance to find the fundamental precision bounds that hold in realistic noisy conditions.

So far the analysis of noisy measurement schemes was limited only to schemes governed by linear generators leading to some very pessimistic results \cite{huelga97, dorner09, kolodynski10, knysh11, escher11, demko12, escher12, knysh13}. Although, under certain circumstances such as the presence of non-Markovian environments \cite{matsuzaki11,chin12}, or frequency estimation with transversal noise \cite{chaves13}, some improvement in the precision with respect to the shot-noise limit can be observed. In our study, we derive the ultimate bounds on the performance of noisy nonlinear measurement schemes and determine whether such schemes can offer any improvement with respect to their noisy linear counterparts. In particular, we investigate the performance of the second-order estimation scheme in the presence of  the two most common types of noise, photon loss and phase diffusion, using a number of different techniques \cite{escher11,toth13,hofmann11,escher12}. While photon loss is typically considered the most dominant source of noise for optical quantum systems, in atomic clocks, for example, phase fluctuations may be more important \cite{borregaard13}.

The paper is organized as follows. In Sec.~\ref{sec::general}, we establish rigorous precision bounds for noisy nonlinear quantum metrology by applying the methods developed in Ref.~\cite{escher11}. Based on this approach, in Secs.~\ref{sec::loss} and ~\ref{sec::diffusion} we derive the ultimate precision bounds for the second-order estimation scheme plagued by photon loss and phase-diffusion noise. Furthermore, in the section on photon loss we make a comparison with other methods used in lossy quantum metrology \cite{toth13, hofmann11} and in the section on phase diffusion we investigate a possible improvement to the performance in a situation when we have some control over the initial state of the environment. We conclude with some final remarks.

\section{General bound on the precision in noisy nonlinear quantum metrology}\label{sec::general}\noindent
In idealized noiseless measurement schemes it is typically assumed that the probe system $S$ evolves under unitary conditions. However, in practice it is impossible to separate the probe system $S$ from its environment $E$, which leads to an unavoidable leakage of information about the parameter $\phi$ to the environment \cite{escher11}. The nonunitary evolution of the probe system in noisy conditions may be described in terms of the Kraus operators $E_{k}$ as
\begin{equation} \label{nlin}
\rho_{S}(\phi) = \sum_{k} E_{k}(\phi) \rho_{S}(0) E^{\dagger}_{k}(\phi)\, ,
\end{equation}
where $\sum_{k} E^{\dagger}_{k}(\phi) E_{k}(\phi) = \mathcal{I}$. We further assume that the Kraus operators act on the constituents of the probe in a nontrivial nonlinear way. Depending on the order of nonlinearity this dynamical evolution can be highly complicated \cite{zwierz10}. For the sake of simplicity, it is  often assumed (although we do not make any such assumptions in this work) that
\begin{equation}
E_{k}(\phi) = E_{k} \, U_{S}(\phi)\, ,
\end{equation}
where $E_{k}$ is a $\phi$-independent Kraus operator that describes only the noise process, which may nevertheless affect the constituents of the probe in some nontrivial nonlinear way, and $U_{S}(\phi) = e^{-i \phi G}$ with $G$ denoting a nonlinear operator generating translations in the parameter $\phi$. In calling  $G$ nonlinear, we mean that it scales nonlinearly with the size of the system. For example, in quantum optics, the parameter shift can be generated by a combination of optical nonlinearities and multiple passes expressed as a nonlinear function of the photon number operators $\hat{n}_{j}$ in each mode:
\begin{equation}
G = \sum^{N}_{j = 1} p_{j} \hat{n}^{q}_{j}\, .
\end{equation}
Here $p_{j}$ denotes the number of passes through the $\phi$-dependent medium experienced by the $j$th mode and $q$ is the order of the nonlinearity \cite{hall12}.

Below, we establish rigorous precision bounds for noisy nonlinear quantum metrology by adopting the methods developed in Ref.~\cite{escher11}. By considering the state of the probe system $S$ together with the state of its environment $E$ we reduce the problem at hand to the problem of parameter estimation governed by a unitary evolution $U_{S,E}(\phi)$. Accordingly, the state of the probe system $S$ initialized in a pure state $|\psi\rangle_{S}$ and its environment $E$ is given by
\begin{equation}\label{kraustate}
|\Psi_{S,E}(\phi) \rangle = U_{S,E}(\phi) |\psi\rangle_{S} |\theta \rangle_{E} = \sum_{k}  \left[ E_{k}(\phi) |\psi\rangle_{S} \right] |k\rangle_{E}\, ,
\end{equation}
where $|\theta \rangle_{E}$ is the initial state of the environment equipped with an orthonormal basis $|k\rangle_{E}$. The above state represents one of many possible purifications of the noisy probe state $\rho_{S}(\phi)$. For a pure state of the probe system and its environment, the quantum Fisher information $F_{Q}[|\Psi_{S,E}(\phi) \rangle]$  is given by \cite{braunstein94,braunstein96}
\begin{equation}\label{QFI}
\frac{F_{Q}[|\Psi_{S,E}(\phi) \rangle]}{4} = \left[ \frac{d \langle \Psi_{S,E}|}{d \phi} \frac{d |\Psi_{S,E}\rangle}{d \phi} - \left| \frac{d \langle \Psi_{S,E}|}{d \phi} |\Psi_{S,E}\rangle \right|^2 \right]\, .
\end{equation}
It was shown in Ref.~\cite{escher11} that $F_{Q}[|\Psi_{S,E}(\phi) \rangle]$ provides an upper bound on the quantum Fisher information $F_{Q}[\rho_{S}(\phi)]$:
\begin{equation}\label{QFIbound}
F_{Q}[\rho_{S}(\phi)] \leq F_{Q}[|\Psi_{S,E}(\phi) \rangle]\, .
\end{equation}
Combining Eqs.~(\ref{CRbound}) and (\ref{QFIbound}) we obtain a lower bound on the square root of the mean-square error of a noisy nonlinear measurement scheme
\begin{equation}
\Delta \phi \geq \frac{1}{\sqrt{m F_{Q}[\rho_{S}(\phi)]}} \geq \frac{1}{\sqrt{m F_{Q}[|\Psi_{S,E}(\phi) \rangle]}}
\end{equation}
with $F_{Q}[|\Psi_{S,E}(\phi) \rangle]$ expressed in terms of the  Kraus operators as
\begin{equation}
F_{Q}[|\Psi_{S,E}(\phi) \rangle] = 4 \left[ \langle H_{1}(\phi) \rangle - \langle H_{2}(\phi) \rangle^{2} \right] ,
\end{equation}
where the averages are calculated in $|\psi\rangle_{S}$ and
\begin{eqnarray}
H_{1}(\phi) &=& \sum_{k} \frac{d E^{\dagger}_{k}(\phi)}{d \phi} \frac{d E_{k}(\phi)}{d \phi}\, , \\
H_{2}(\phi) &=& \sum_{k} \frac{d E^{\dagger}_{k}(\phi)}{d \phi} E_{k}(\phi)\, .
\end{eqnarray}
The bound in Eq.~(\ref{QFIbound}) is saturated by minimizing over all possible purifications of $\rho_{S}(\phi)$ \cite{escher11}. That is,
\begin{equation}\label{minimum}
F_{Q}[\rho_{S}(\phi)] = \min_{\{E_{k}(\phi)\}} F_{Q}[|\Psi_{S,E}(\phi) \rangle]\, .
\end{equation}
Let us emphasize that this minimization is taken over all sets $\{E_{k}(\phi)\}$ that leave the map in Eq.~(\ref{nlin}) unchanged. We now apply the above approach to the second-order phase estimation scheme plagued by photon loss and phase diffusion.

\section{Photon loss}\label{sec::loss}\noindent
The damaging effect of photon loss on the performance of linear estimation schemes has been extensively studied in recent years \cite{dorner09, kolodynski10, kacprowicz10, knysh11}. It has been repeatedly shown that in the presence of photon loss the precision scaling offered by linear estimation schemes deteriorates to the classical limit. Specifically, in the asymptotic limit of large $N$, the presence of loss unavoidably leads to the shot-noise-like scaling $c/\sqrt{N}$, where the improvement over the classical measurement is limited to a constant factor $c$ independent of $N$ \cite{kolodynski10, knysh11}. In the following, we derive the ultimate precision bounds for the second-order estimation scheme in the presence of photon loss and determine whether a nonlinear scheme offers any improvement with respect to its linear counterpart.

We consider a typical optical interferometric setup: an optical probe system $S$ prepared in some initial pure state given by
\begin{equation}
\rho_{S}(0) = |\psi\rangle_{S}\langle \psi| = \sum_{n, m = 0}^{\infty} \rho_{nm}(0) |n \rangle \langle m|\,
\end{equation}
is fed into a two-arm Mach-Zehnder interferometer. The phase shift $\phi$ is introduced with a dispersive second-order Kerr medium placed in one of the interferometric arms. That is, the state of the probe system is evolved under a unitary operator $U_{S}(\phi) = e^{-i \phi \hat{n}^{2}}$. The presence of photon loss is described with the set of Kraus operators:
\begin{equation}\label{onlyKraus}
E_{k} = \sqrt{\frac{(1-\eta)^k}{k!}} \eta^{\frac{\hat{n}}{2}} \hat{a}^k\, ,
\end{equation}
where $\eta$ quantifies the strength of photon loss with $\eta = 0$ and $\eta = 1$ corresponding to the complete absorption and lossless regimes, respectively. Schematically, the process of photon loss is modeled with a beam splitter characterized with transmissivity $\eta$ located in the dispersive arm of the interferometer. The beam splitter reflects each photon with probability $1-\eta$; hence, we consider this to be a \textit{linear} form of photon loss. We limit the discussion in this paper to the case of a one-arm photon loss, however, it should be relatively straightforward to extend our results to the case of a two-arm photon loss.

\begin{figure}[!t]
\centering
\includegraphics[width=7.8cm]{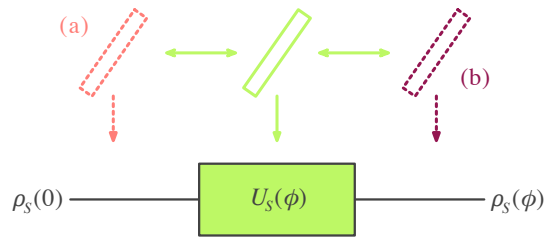}
\caption{Schematic illustrating various photon loss scenarios. The dashed beam splitters model photon loss occurring prior to (a) or after (b) the interaction with the dispersive Kerr medium. The solid beam splitter models photon loss incorporated within the evolution. \label{scenarios}}
\end{figure}

\subsection{Photon loss scenarios}
Depending on the location of the beam splitter the method used here leads to distinct precision bounds. We begin by examining two standard photon loss scenarios \cite{escher11}. First, we consider a situation where photon loss occurs after the interaction with the dispersive Kerr medium [see Fig.~\ref{scenarios}(b)]. Using Eqs.~(\ref{QFI}) and (\ref{onlyKraus}) we obtain a trivial lossless upper bound on the quantum Fisher information $F_{Q}[|\Psi_{S,E}(\phi) \rangle] = 4 \left(\Delta \hat{n}^2\right)^{2}_{|\psi\rangle_{S}}$, where $\left(\Delta \hat{n}^2\right)^{2}_{|\psi\rangle_{S}}$ is the variance of the squared photon number operator calculated in the initial probe state $|\psi\rangle_{S}$. In this scenario, $F_{Q}[|\Psi_{S,E}(\phi) \rangle]$ will typically scale as $N^4$. Second, in the situation where photon loss takes place prior to the interaction with the dispersive Kerr medium [see Fig.~\ref{scenarios}(a)], we obtain a slightly stronger bound. In this case, the upper bound on $F_{Q}[|\Psi_{S,E}(\phi) \rangle]$ is given by
\begin{eqnarray}\label{beforebound}
F_{Q}[|\Psi_{S,E}(\phi) \rangle] &=& 4 \left[ \eta^{4} \left(\Delta \hat{n}^{2}\right)^{2}_{|\psi\rangle_{S}} - 6\eta^{3} (\eta - 1) \langle \hat{n}^{3} \rangle_{|\psi\rangle_{S}} \right. \nonumber \\
  &+& \eta^{2} (11 \eta^{2} - 18 \eta + 7) \langle \hat{n}^{2} \rangle_{|\psi\rangle_{S}} \nonumber \\
  &-& \eta (6 \eta^{3} - 12 \eta^{2} + 7 \eta - 1) \langle \hat{n} \rangle_{|\psi\rangle_{S}} \nonumber \\
  &+& 2 \eta^{3} (\eta - 1) \langle \hat{n}^{2} \rangle_{|\psi\rangle_{S}} \langle \hat{n} \rangle_{|\psi\rangle_{S}} \nonumber \\
  &-& \left. \eta^{2} (\eta - 1)^{2} \langle \hat{n} \rangle^{2}_{|\psi\rangle_{S}} \right]\, .
\end{eqnarray}
This scales as $N^4_\text{eff}$, where $N_\text{eff} = \eta N$ corresponds to the effective average number of photons that survive the loss process and are subsequently used in the measurement. In order to derive an upper bound on $F_{Q}[|\Psi_{S,E}(\phi) \rangle]$ which shows the difference in scaling expected from loss, we need to incorporate the loss within the evolution \cite{escher11}.

Guided by the above observations and the form of the commutation relation between $U_{S}(\phi)$ and $E_{k}$ we suggest the following set of the generalized Kraus operators to model the presence of photon loss in a nonlinear (quadratic) optical phase shift material
\begin{equation}\label{nonKraus}
E_{k}(\phi) = \sqrt{\frac{(1-\eta)^k}{k!}} \eta^{\frac{\hat{n}}{2}} \hat{a}^k e^{-i \phi(\hat{n}^2 - 2\lambda_{1} k \hat{n} + \lambda_{2} k^2)}\, .
\end{equation}
Here $\lambda_{1}$ and $\lambda_{2}$ are the variational parameters with $\lambda_{1} = \lambda_{2} = 0$ ($\lambda_{1} = \lambda_{2} = 1$) corresponding to photon loss occurring after (before) the dispersive Kerr medium. The variational parameters allow us to minimize the value of the quantum Fisher information $F_{Q}[|\Psi_{S,E}(\phi) \rangle]$, for a fixed $\eta$, which in turn gives us the strongest lower bound on the precision \cite{escher11}.

As part of the optimization process, we first analyze the structure of $E_{k}(\phi)$. We note that the variational parameters $\lambda_{1}$ and $\lambda_{2}$ play very different roles. By varying $\lambda_{1}$ we can interpolate between two different physical regimes of photon loss occurring before or after the dispersive Kerr medium. This interpolation is described by a family of generalized Kraus operators. In order to limit the following minimization to the physically relevant family of generalized Kraus operators we have to impose the following constraint $0 \leq \lambda_{1} \leq 1$. By varying $\lambda_{2}$ we are not leaving this family of Kraus operators as we are not changing the map defined in Eq.~(\ref{nlin}) for a given member of the family. Therefore, when minimizing the quantum Fisher information $F_{Q}[|\Psi_{S,E}(\phi) \rangle]$ we can vary $\lambda_{2}$ without a constraint. Given Eqs.~(\ref{QFI}) and (\ref{nonKraus}) we can write
\begin{equation}\label{lossyfisher}
F_{Q}[|\Psi_{S,E}(\phi) \rangle] = 4 \left[\langle H_{1}(\phi) \rangle_{|\psi\rangle_{S}} - \langle H_{2}(\phi) \rangle^{2}_{|\psi\rangle_{S}} \rangle \right],
\end{equation}
where
\begin{eqnarray}
\langle H_{1}(\phi) \rangle_{|\psi\rangle_{S}} &=& \sum_{n=0}^{\infty} \rho_{nn}(0) \nonumber \\
&\times& \sum_{k=0}^{\infty} c_{nk}(\eta) \left( n^{2} - 2 \lambda_{1} k n + \lambda_{2} k^{2} \right)^{2}, \\
\langle H_{2}(\phi) \rangle_{|\psi\rangle_{S}} &=& \sum_{n=0}^{\infty} \rho_{nn}(0) \nonumber \\
&\times& \sum_{k=0}^{\infty} c_{nk}(\eta) \left( n^{2} - 2 \lambda_{1} k n + \lambda_{2} k^{2}\right)
\end{eqnarray}
with
\begin{equation}
c_{nk}(\eta) = \frac{(1-\eta)^k}{k!} \eta^{n-k} \frac{n!}{(n-k)!}\, .
\end{equation}
Since any general analytical minimization of $F_{Q}[|\Psi_{S,E}(\phi) \rangle]$ given in Eq.~(\ref{lossyfisher}) over both $\lambda_{1}$ and $\lambda_{2}$ with a constraint $0 \leq \lambda_{1} \leq 1$ proved to be very hard we proceed by expressing the first four moments of the number operator in terms of the average number of photons $N$ [$\langle \hat{n} \rangle_{|\psi\rangle_{S}} = N$] present in the input probe state. For one-mode Gaussian states we use the following upper bounds
\begin{eqnarray}
\langle \hat{n}^4 \rangle_\text{Gauss} &\leq& 105 N^4 + 180 N^3 + 84 N^2 + 8 N\, , \label{sv4} \\
\langle \hat{n}^3 \rangle_\text{Gauss} &\leq& 15 N^3 + 18 N^2 + 4 N\, , \label{sv3} \\
\langle \hat{n}^2 \rangle_\text{Gauss} &\leq& 3 N^2 + 2 N \label{sv2}
\end{eqnarray}
that are saturated by squeezed vacuum states \cite{monras06}. Then, we perform the numerical minimization of $F_{Q}[|\Psi_{S,E}(\phi) \rangle]$ over $\lambda_{1}$ and $\lambda_{2}$ with a constraint $0 \leq \lambda_{1} \leq 1$. We find, in particular, that the minimum of $F_{Q}[|\Psi_{S,E}(\phi) \rangle]$ is always attained for $\lambda_{1} = 1$. Given this bit of information we obtain an analytic minimum $F^\text{min}_{Q}[|\Psi_{S,E}(\phi) \rangle]$ by minimizing $F_{Q}[|\Psi_{S,E}(\phi) \rangle]$ over $\lambda_{2}$ with $\lambda_{1} = 1$. Sadly, this minimum takes a very convoluted form [for details see the Appendix]. Therefore, for the sake of clarity we present here its asymptotic form that holds in the limit of large $N$
\begin{equation}
F_{Q}^\text{min}[|\Psi_{S,E}(\phi) \rangle] \leq \frac{\eta}{1 - \eta} \, 240 \, \eta^{2} N^{3} + \mathcal{O}\left[N^{2}\right]\,
\end{equation}
and leads to the ultimate asymptotic lower bound on the error of the second-order estimation scheme affected by a one-arm photon loss valid for any input state
\begin{eqnarray}
\Delta \phi_{\rm LB}^{N \rightarrow \infty} &\geq& \sqrt{\frac{1}{F^\text{min}_{Q}[|\Psi_{S,E}(\phi) \rangle]}} \nonumber \\
&\geq& \sqrt{\frac{1 - \eta}{\eta}} \frac{1}{4 \eta \sqrt{15} N^{3/2}} + \mathcal{O}\left[\frac{1}{N^{5/2}}\right] . \label{ultimate_asymptotic_loss}
\end{eqnarray}
We conclude that in the presence of photon loss the precision scaling is reduced to the classical shot-noise-like limit $1/N^{3/2}$ of the second-order estimation scheme \cite{boixo08}. This result is reminiscent of analogous results in lossy linear estimation, where a $1/N$ scaling deteriorates to \cite{kolodynski10, knysh11}
\begin{equation}
\Delta \phi_{\rm LB}^{N \rightarrow \infty} \geq \sqrt{\frac{1 - \eta}{\eta}} \frac{1}{2 N^{1/2}}\, .
\end{equation}
Therefore, it appears that the nonlinear estimation schemes are not the solution to the problem of photon loss in optical interferometry, that is, those schemes \textit{are not} more resistant to photon loss than their linear counterparts.

\subsection{Comparison}
For the sake of completeness, we compare the above precision bounds with two further precision bounds that can be obtained in the following way. First, in the situation where photon loss takes place prior to the interaction with the dispersive Kerr medium [see Fig. 1(a)] the state of the probe system immediately before the phase shift can be written as
\begin{equation}
\rho_{S} = \sum_{k} p_{k} |\psi_{k} \rangle_{S} \langle \psi_{k}| \ \ \mbox{with} \ \ |\psi_{k} \rangle_{S} = \frac{E_{k} |\psi \rangle_{S}}{\sqrt{p_{k}}}\, ,
\end{equation}
where $p_{k}$ is the probability of losing $k$ photons defined as $p_{k} = \mbox{Tr}\left[E_{k} |\psi \rangle_{S}\langle \psi| E^{\dagger}_{k}\right]$.
Due to the convexity of the quantum Fisher information \cite{toth13} we can obtain the averaged upper bound
\begin{equation}\label{averagedbound}
F_{Q}[\rho_{S}(\phi)] \leq \sum_{k} p_{k} \, F_{Q}[|\psi_{k} \rangle_{S}]\, ,
\end{equation}
where $F_{Q}[|\psi_{k} \rangle_{S}] = 4 \left\langle \left(\hat{n}^{2} - \langle \hat{n}^{2} \rangle_{|\psi_{k}\rangle_{S}}\right)^{2} \right\rangle_{|\psi_{k}\rangle_{S}}$.\\
Second, in the situation where photon loss occurs after the interaction with the dispersive Kerr medium [see Fig.~\ref{scenarios}(b)] another upper bound on $F_{Q}[\rho_{S}(\phi)]$ can be obtained from the statistics of weak values \cite{hofmann11}:
\begin{equation}\label{weakbound}
F_{Q}[\rho_{S}(\phi)] \leq \sum_{k} p_{k} \, F_{Q}[_{\Pi_{k}}\langle \hat{n}^{2} \rangle_{|\psi \rangle_{S}}]\, ,
\end{equation}
where $F_{Q}[_{\Pi_{k}}\langle \hat{n}^{2} \rangle_{|\psi \rangle_{S}}] = 4 \left\langle \left(\hat{n}^{2} - \, _{\Pi_{k}}\langle \hat{n}^{2} \rangle_{|\psi \rangle_{S}}\right)^{2} \right\rangle_{|\psi \rangle_{S}}$ and
\begin{equation}
_{\Pi_{k}}\langle \hat{n}^{2} \rangle_{|\psi \rangle_{S}} = \mbox{Re} \left(\frac{\mbox{Tr}[\Pi_{k} \, \hat{n}^{2} \, |\psi \rangle_{S}\langle \psi|]}{\mbox{Tr}[\Pi_{k} \, |\psi \rangle_{S}\langle \psi|]} \right)\, ,
\end{equation}
where $\Pi_{k}=E^{\dagger}_{k} E_{k} $  is an element of the Positive Operator-Valued Measure (POVM) $\{\Pi_{k}\}$ that effectively describes a lossy measurement.

\begin{figure}[!t]
\centering
\includegraphics[scale=0.9]{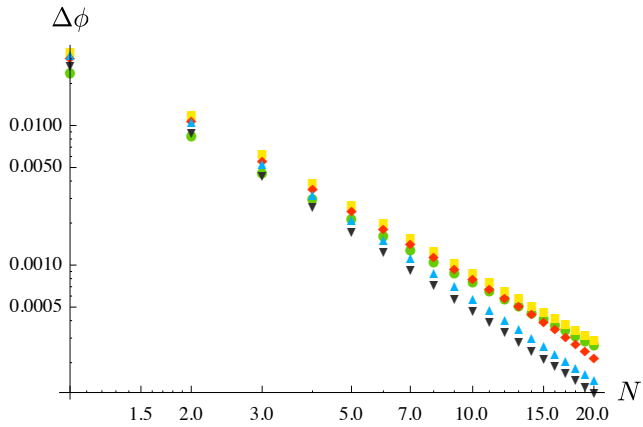}
\caption{The log-log dependence of various precision bounds on the average number of photons $N$ for a squeezed vacuum state plotted for $\eta = 0.90$.
 From the top to the bottom the curves correspond to (1) the ultimate \textit{analytic} bound given in Eq.~(\ref{analyticbound}) [with $F^\text{min}_{Q}[|\Psi_{S,E}(\phi) \rangle]$ defined in Eq.~(\ref{Fmin_sv})], which is indistinguishable from the averaged bound given in Eq.~(\ref{averagedbound}) (yellow squares); (2) the ultimate \textit{asymptotic} bound given in  Eq.~(\ref{ultimate_asymptotic_loss}) (green dots); (3) the weak-valued bound given in  Eq.~(\ref{weakbound}) (red rhombi/diamonds); (4) the bound corresponding to photon loss occurring prior to the interaction with the dispersive Kerr medium given in Eq.~(\ref{beforebound}) (blue triangles); (5) the trivial lossless bound corresponding to photon loss occurring after the interaction with the dispersive Kerr medium (black inverted triangles). \label{variousbounds}}
\end{figure}

The weak-valued precision bound is plotted in Fig.~\ref{variousbounds} together with the other precision bounds discussed in Sec.~\ref{sec::loss}~A. Due to the numerical issues the weak-valued bound is plotted in its approximated form with photon loss limited to up to 30 photons, that is, $k_\text{max} = 30$. However, this does not invalidate the corresponding curve as even for $N = 20$ the approximation accounts for almost all loss, that is, $\sum^{30}_{k = 0} p_{k} = 0.9998$. We observe that the weak-valued bound, while quite tight for small $N$, appears to scale no better than the trivial lossless bound and the bound of Eq.~(\ref{beforebound}) in the asymptotic limit. Instead of the averaged precision bound we decided to plot the full analytic bound given in the Appendix [see Eq.~(\ref{analyticbound}) with $F^\text{min}_{Q}[|\Psi_{S,E}(\phi) \rangle]$ defined in Eq.~(\ref{Fmin_sv})] because we found that these bounds are indistinguishable from each other for all values of $N$ up to 20.

\section{Non-dissipative phase-diffusion noise}\label{sec::diffusion}\noindent
The influence of phase diffusion on the performance of linear estimation schemes was recently studied in Refs.~\cite{brivio10, genoni11, escher12}. In Ref.~\cite{escher12} it was shown that the ultimate precision bound for a linear estimation scheme affected by phase diffusion contains a constant term independent of the average number of photons $N$. Hence, it is believed that the presence of phase diffusion is even more detrimental to the performance of linear estimation than the presence of photon loss. Can the second-order estimation scheme offer any improvement or does its performance deteriorate to a similar constant bound? In the following, we address this question using the variational approach to the method of purifications introduced in Ref.~\cite{escher12}.

We model the phase-diffusion noise as arising from the interaction between the probe system $S$ and the environment $E$, represented  by uncertainty in the position of one of the interferometric mirrors. This gives rise to \textit{linear} phase-diffusion noise with the corresponding interaction generated by $\hat{n} \hat{x}_{E}$, where $\hat{x}_{E}$ is the position operator of the mirror. We note that a different physical model could give rise to a \textit{second-order} phase-diffusion noise with the corresponding interaction proportional to $\hat{n}^2 \hat{x}_{E}$, and we return to this at the end of this section.

Given the above physical model, the state of the probe system $S$ and its environment $E$ can be written as
\begin{equation}\label{purification}
|\Phi_{S,E}(\phi)\rangle = e^{-i \phi \hat{n}^2} e^{i 2 \beta \hat{n} \hat{x}_{E}} |\psi\rangle_{S} |\theta \rangle_{E}\, ,
\end{equation}
where $\beta$ denotes the strength of phase-diffusion noise \cite{escher12} and $|\theta \rangle_{E}$ denotes the initial Gaussian state of the environment which in the position eigenbasis reads
\begin{equation}
|\theta \rangle_{E} = \int_{-\infty}^{\infty} \frac{d x_{E}}{\sqrt[4]{\pi \Delta^2}} \exp\left[{-\frac{x^{2}_{E}}{2 \Delta^2}}\right] |x_{E}\rangle
\end{equation}
with $\Delta$ being the standard deviation of $x_{E}$. It is easy to verify that the state given in Eq.~(\ref{purification}) represents one of many allowed purifications of the noisy probe system $\rho_{S}(\phi)$, that is, the probe system considered alone without any knowledge about its environment is in the state
\begin{eqnarray}
\rho_{S}(\phi) &=& \text{Tr}_{E}[|\Phi_{S,E}(\phi)\rangle \langle\Phi_{S,E}(\phi)|] \nonumber \\
&=& \sum_{n, m = 0}^{\infty} \rho_{nm}(0) e^{-i \phi (n^2 - m^2) - \beta^2 \Delta^2 (n - m)^2} |n \rangle \langle m|\, . \nonumber \\
\end{eqnarray}
It is also straightforward to verify that this particular purification leads to a trivial noiseless upper bound on the quantum Fisher information $F_{Q}[|\Psi_{S,E}(\phi) \rangle] = 4 \left(\Delta \hat{n}^2\right)^{2}_{|\psi\rangle_{S}}$. That is because this purification allows one to find out (after the interaction) the exact position $x$ of the mirror, which would allow one to correct for the phase shift $2\beta x$ it introduced. A stronger upper bound can be determined by minimizing $F_{Q}[|\Psi_{S,E}(\phi) \rangle]$ over all purifications of $\rho_{S}(\phi)$.

\subsection{Variational method}
It was shown in Ref.~\cite{escher12} that the optimal purification of $\rho_{S}(\phi)$ that minimizes $F_{Q}[|\Psi_{S,E}(\phi) \rangle]$ can be written as
\begin{equation}
|\Psi_{S,E}(\phi)\rangle = U_{E}(\phi) |\Phi_{S,E}(\phi)\rangle
\end{equation}
with $U_{E}(\phi) = e^{-i \phi \hat{h}_\text{min}}$ being a unitary operator acting only on the environment generated by the optimal Hermitian operator $\hat{h}_\text{min}$, which is implicitly given by
\begin{equation}\label{eqhopt}
\frac{1}{2} \left(\hat{h}_\text{min} \rho_{E}(\phi) + \rho_{E}(\phi) \hat{h}_\text{min}\right) = \text{Tr}_{S} \{ \mathcal{D} \left[ \rho_{S,E}(\phi) \right] \}\, ,
\end{equation}
where $\rho_{E}(\phi)$ is the reduced state of the environment, and $\mathcal{D} \left[ \rho_{S, E}(\phi) \right]$ is defined as
\begin{equation}
\mathcal{D} \left[ \rho_{S, E}(\phi) \right] \equiv \frac{1}{2 i}\left[ \frac{d |\Phi_{S,E}\rangle}{d\phi}\langle \Phi_{S,E}| - | \Phi_{S,E}\rangle \frac{d \langle\Phi_{S,E}|}{d\phi} \right]\, .
\end{equation}
Based on the initial purification given in Eq.~(\ref{purification}) the right-hand side of Eq.~(\ref{eqhopt}) evaluates to
\begin{equation}
\text{Tr}_{S} \{ \mathcal{D} \left[ \rho_{S,E}(\phi) \right]\} = \frac{1}{2} \left(\hat{O} \rho_{E}(\phi) + \rho_{E}(\phi) \hat{O}^{\dagger}\right)\, ,
\end{equation}
where the operator $\hat{O}$ is given by
\begin{equation}
\hat{O} = \frac{\hat{I}_{E}}{4 \beta^2 \Delta^2} + \frac{\hat{x}^{2}_{E}}{4 \beta^2 \Delta^4} + \frac{i \hat{x}_{E} \hat{p}_{E}}{2 \beta^2 \Delta^2} - \frac{\hat{p}^{2}_{E}}{4 \beta^2}\, .
\end{equation}
Alas, $\hat{O}$ is not Hermitian. However, we can be guided by the form of $\hat O$ to guess an operator $\hat h_{\rm ansatz}$, which will give a useful bound. (Note that \textit{any} choice of Hermitian $\hat h_{\rm ansatz}$ will give a valid bound; the challenge is to make a choice that results in an upper bound on $F_{Q}[|\Psi_{S,E}(\phi) \rangle]$ that is stronger than the bound introduced earlier.)
To obtain $\hat h_{\rm ansatz}$ we discard the third non-Hermitian term [and also the first term, which only contributes an overall phase factor to the optimal purification $|\Psi_{S,E}(\phi)\rangle$]. This gives
\begin{equation}
\hat{h}_\text{ansatz} = \frac{\hat{x}^{2}_{E}}{4 \beta^2 \Delta^4} - \frac{\hat{p}^{2}_{E}}{4 \beta^2}
\end{equation}
which suggests the following  variational form of the unitary operator $U_{E}(\phi)$
\begin{equation}
U_{E}(\phi; \lambda_{1}, \lambda_{2}) = \exp \left[- i \phi \left( \lambda_{1} \frac{\hat{x}^{2}_{E}}{4 \beta^2 \Delta^4} - \lambda_{2} \frac{\hat{p}^{2}_{E}}{4 \beta^2} \right) \right]\, .
\end{equation}
We allow here for two variational parameters $\lambda_{1}$ and $\lambda_{2}$ so that we can derive the strongest possible upper bound on $F_{Q}[|\Psi_{S,E}(\phi) \rangle]$.

The presence of two non-commuting terms in $U_{E}(\phi; \lambda_{1}, \lambda_{2})$ produces a fairly complicated purification:
\begin{eqnarray}\label{optimalpure}
|\Psi_{S,E}(\phi)\rangle &=& N_{c} \int_{-\infty}^{\infty} d x_{E} \exp\left[{-\frac{x^{2}_{E}}{2 \Delta^2}} + i 2 \beta \mathcal{C} \hat{n} x_{E}\right] |x_{E} \rangle \nonumber \\
&\times& e^{-i \phi \hat{n}^2} |\psi\rangle_{S} \, ,
\end{eqnarray}
where $N_{c}$ is the normalization constant and $\mathcal{C} = \exp(-i \xfrac{\sqrt{\lambda^2 \phi^2}}{2 \beta^2 \Delta^2})$. The above purification holds only for $\lambda_{2} = - \lambda_{1} = \lambda$. Naturally, one can derive a general purification holding for arbitrary $\lambda_{1}$ and $\lambda_{2}$, however, this purification is even more convoluted than the one given in Eq.~(\ref{optimalpure}). Since any further analytical derivation proved to be intractable, we simplify the problem by assuming that
\begin{equation}\label{range}
|\phi| \ll \frac{2 \beta^2 \Delta^2}{|\lambda|}\, .
\end{equation}
This restriction will become important later.

Given the purification in Eq.~(\ref{optimalpure}) and the approximation of small $|\phi|$ we obtain an analytical upper bound on the quantum Fisher information $F_{Q}[|\Psi_{S,E}(\phi) \rangle]$
\begin{equation}
F_{Q}[|\Psi_{S,E}(\phi) \rangle] = (1 - \lambda)^2 4 \left(\Delta \hat{n}^2\right)^{2}_{|\psi\rangle_{S}} + \frac{\lambda^2}{2 \beta^2 \Delta^2} 4 \langle \hat{n}^2 \rangle_{|\psi\rangle_{S}}\, ,
\end{equation}
which is minimized for $\lambda$ given by
\begin{equation}\label{optimallam}
\lambda_\text{min} = \frac{2 \beta^2 \Delta^2 \left(\Delta \hat{n}^2\right)^{2}_{|\psi\rangle_{S}}}{\langle \hat{n}^2 \rangle_{|\psi\rangle_{S}} + 2 \beta^2 \Delta^2 \left(\Delta \hat{n}^2\right)^{2}_{|\psi\rangle_{S}}}\, .
\end{equation}
We note that for $\lambda = 0$ we recover the upper bound on $F_{Q}[|\Psi_{S,E}(\phi) \rangle]$ in the noiseless case. Finally, we can state the main result of this section. A lower bound on the error of the second-order estimation scheme in the presence of a linear phase-diffusion noise is given by
\begin{eqnarray}\label{phasediff}
\Delta \phi_{\rm LPD} &\geq& \sqrt{\frac{1}{F^\text{min}_{Q}[|\Psi_{S,E}(\phi) \rangle]}} \nonumber \\
&=& \sqrt{\frac{1}{4 \left(\Delta \hat{n}^2\right)^{2}_{|\psi\rangle_{S}}} + \frac{2 \beta^2 \Delta^2}{4 \langle \hat{n}^2 \rangle_{|\psi\rangle_{S}}}}\, .
\end{eqnarray}
Given the upper bounds
\begin{eqnarray} \label{gauss_bounds1}
\left(\Delta \hat{n}^2\right)_\text{Gauss}^2 &\leq& 96 N^4 + 168 N^3 + 80 N^2 + 8 N\, , \\
\langle \hat{n}^2 \rangle_\text{Gauss} &\leq& 3 N^{2} + 2 N \label{gauss_bounds2}
\end{eqnarray}
that are saturated by squeezed vacuum states \cite{monras06}, the asymptotic limit of large $N$ of the ultimate bound given in Eq.~(\ref{phasediff}) can be written as
\begin{equation}\label{asymptotic_phase}
\Delta \phi_{\rm LPD}^{N \rightarrow \infty} \geq \frac{\beta \Delta}{\sqrt{6} N} + \mathcal{O}\left[\frac{1}{N^2}\right]\, .
\end{equation}
Interestingly, for coherent states, for which
\begin{eqnarray}\label{coh_bounds1}
\left(\Delta \hat{n}^2\right)_\text{coh}^2 &=& 4 N^3 + 6 N^2 + N\, , \\
\langle \hat{n}^2 \rangle_\text{coh} &=& N^2 + N \label{coh_bounds2}
\end{eqnarray}
this scaling for the phase uncertainty differs only by a $\sqrt{3}$ multiplicative factor. That is, the scaling reads
\begin{equation}\label{asymptotic_phase_coherent}
\Delta \phi_{\rm LPD}^{N \rightarrow \infty} \geq \frac{\beta \Delta}{\sqrt{2} N} + \mathcal{O}\left[\frac{1}{N^2}\right]\, .
\end{equation}
We observe an important qualitative difference between our bounds and the lower bounds on the precision of the linear estimation scheme given in Ref.~\cite{escher12}. Our result contains no constant terms and thus retains its dependence on the average photon number $N$. Hence, the error of the second-order estimation scheme $\Delta \phi_{\rm LPD}^{N \rightarrow \infty}$ can go to zero as $N$ goes to infinity. This result proves that in the presence of a linear phase-diffusion noise the second-order estimation scheme is superior to its linear counterpart.

\subsection{Environment}
The asymptotic bounds given in  Eqs.~(\ref{asymptotic_phase}) and (\ref{asymptotic_phase_coherent}) depend on the product of the strength of the coupling $\beta$ and the position variance $\Delta$  in the initial state of the environment (mirror). Thus by creating a highly squeezed initial state of the environment, the error could be made arbitrarily small. However, we recall that this result holds only for a particular range of $\phi$ around zero. By combining Eq.~(\ref{range}) with $|\lambda| = \lambda_\text{min}$ we obtain
\begin{equation}\label{limitation}
|\phi| \ll 2 \beta^2 \Delta^2 + \frac{\langle \hat{n}^2 \rangle_{|\psi\rangle_{S}}}{\left(\Delta \hat{n}^2\right)^{2}_{|\psi\rangle_{S}}}\, .
\end{equation}
Therefore, for a given \textit{optimally} squeezed input probe state [see Eqs.~(\ref{gauss_bounds1}) and (\ref{gauss_bounds2})] and a finite value of $\beta$, an arbitrarily small $\Delta$ implies that the bound in Eq.~(\ref{asymptotic_phase}) may apply only to an arbitrarily small interval of phases. We note, however, that this limitation does not apply to the coherent states, because the interval of phases and the estimation error contain the same linear scaling with respect to $N$ [see Eqs.~(\ref{coh_bounds1}) and (\ref{coh_bounds2})]. Hence, in the case of coherent input probe states, this sort of protection against the phase-diffusion noise introduced through the squeezing of the environment is indeed possible.

It is instructive to examine how the presence of a squeezed environment changes the scaling of the precision bound given in Eq.~(\ref{asymptotic_phase_coherent}).  Without loss of generality (since only the product $\Delta \beta$ appears in the above) we can take $\Delta = 1$ to correspond to the vacuum-state of the mirror motion.  Then we can express the squeezing parameter $\Delta$ in terms of the average number of excitations $N_{E} = {}_{E}\langle \theta |\hat{n} | \theta \rangle_{E} $ present in the environment
(mirror) as
\begin{equation}\label{delta}
\Delta = \exp\left[- \mbox{arcsinh}\sqrt{N_{E}}\right]
\end{equation}
which yields
\begin{equation}
\Delta \phi_{\rm LPD}^{N, N_{E} \rightarrow \infty} \geq \frac{\beta}{2 \sqrt{2} N N^{1/2}_{E}} + \mathcal{O}\left[\frac{1}{N^2}, \frac{1}{N^{3/2}_{E}}\right]\, .
\end{equation}
Thus we can improve the performance of the second-order estimation scheme employing the coherent input probe states in the presence of linear phase-diffusion noise by squeezing the environment. However, this does not change the scaling of the bound with respect to the number of quanta in the probe beam. Moreover, we observe that the scaling with the number of quanta in the environment is only the usual (linear) shot-noise limit, as expected since we consider here a linear phase-diffusion noise.

\subsection{Second-order phase-diffusion noise}
We also briefly note that the lower bounds on the error of the second-order estimation scheme in the presence of \textit{second-order} phase-diffusion noise qualitatively coincide with the lower bounds derived in Ref.~\cite{escher12}:
\begin{eqnarray}
\Delta \phi_{\rm SPD} &\geq& \sqrt{\frac{1}{4 \left(\Delta \hat{n}^2\right)^{2}_{|\psi\rangle_{S}}} + 2 \gamma^2 \Delta^2}\, , \\
\Delta \phi_{\rm SPD}^{N \rightarrow \infty} &\geq& \sqrt{2} \gamma \Delta + \mathcal{O}\left[\frac{1}{N^2}\right]\, ,
\end{eqnarray}
where $\gamma$ denotes the strength of the second-order phase-diffusion noise. This bound holds for any input probe state and for any value of $\phi$.   Interestingly, in this scenario the protection against the phase-diffusion noise introduced above is   possible for any input probe state. The corresponding bounds are
\begin{eqnarray}
&& \Delta \phi_{\rm SPD} \geq \sqrt{\frac{1}{4 \left(\Delta \hat{n}^2\right)^{2}_{|\psi\rangle_{S}}} + \frac{2 \gamma^2}{\exp\left[2 \, \mbox{arcsinh}\sqrt{N_{E}}\right]}}\, , \\
&& \Delta \phi_{\rm SPD}^{N, N_{E} \rightarrow \infty} \geq \frac{\gamma}{\sqrt{2} N^{1/2}_{E}} + \mathcal{O}\left[\frac{1}{N^2}, \frac{1}{N^{3/2}_{E}}\right]\, .
\end{eqnarray}
We note that here the asymptotic bound again scales with the shot-noise-limited precision in terms of $N_E$. In this case, the dominant term in this asymptotic limit is determined fully from the environment. As above, we can conclude that by squeezing the initial state of the environment it is possible to improve the performance of the second-order estimation scheme in the presence of the second-order phase-diffusion noise. By analogy the same result and conclusion holds for the linear estimation scheme affected by the linear phase-diffusion noise.

Naturally, squeezing the motional state of the interferometric mirror is probably not the most practical solution to the problem of phase diffusion. However, we believe that other metrological setups based, for example, on the optomechanical systems, could be better suited for these ideas.

\renewcommand{\arraystretch}{2.0}
\begin{table*}[!ht]
\centering
\begin{tabular}{ c | >{\centering\arraybackslash}m{3.5cm} | >{\centering\arraybackslash}m{3.5cm} >{\centering\arraybackslash}m{0cm} }
  \hline
  \hline
   & linear estimation & second-order estimation \\
  \hline
  photon loss & $\displaystyle \sqrt{\frac{1 - \eta}{\eta}} \frac{1}{2 N^{1/2}}$ & $\displaystyle \sqrt{\frac{1 - \eta}{\eta}} \frac{1}{4 \eta \sqrt{15} N^{3/2}}$ & \\ [2ex]
  \hline
  linear phase diffusion & $\displaystyle \sqrt{2} \beta \Delta \, \left[\mbox{or} \, \frac{\beta}{\sqrt{2} N^{1/2}_{E}}\right]$ & $\displaystyle \frac{\beta \Delta}{\sqrt{2} N} \,  \left[ \mbox{or} \, \frac{\beta}{2 \sqrt{2} N N^{1/2}_{E}} \right]$ & \\ [2ex]
  \hline
  second-order phase diffusion & -- & $\displaystyle \sqrt{2} \gamma \Delta \, \left[ \mbox{or} \, \frac{\gamma}{\sqrt{2} N^{1/2}_{E}} \right]$ & \\ [2ex]
  \hline
  \hline
\end{tabular}
\centering
\caption{Precision bounds for the linear and second-order estimation schemes affected by photon loss and phase diffusion. The precision bound for the second-order estimation scheme in the presence of linear phase diffusion holds for coherent states. \label{summary}}
\end{table*}

\section{Conclusions}\label{sec::conclusions}\noindent
The presence of noise renders the Heisenberg scaling of quantum metrology unachievable asymptotically. We confirm this by providing the ultimate precision bounds for the second-order estimation scheme affected by photon loss and phase diffusion. For a summary of all the main findings see Table~\ref{summary}. We observe that for both types of noise the gain in the phase sensitivity with respect to the linear estimation scheme is given by a multiplicative term $\mathcal{O}(1/N)$, or $\mathcal{O}(1/N^{q-1})$ in the general case of higher-order schemes. This result makes an intuitive sense as one can linearize the number operator $\hat{n}$ around the mean, so that $\hat{n} = N + \delta \hat{n}$ in the linear case, and $\hat{n}^q = N^q + q N^{q-1} \delta \hat{n}$ in the nonlinear case which implies that the relevant gain in the phase sensitivity should scale as $N^{q-1}$.

Regardless of the above similarities, the presence of phase diffusion can be considered more detrimental than the presence of photon loss. Nevertheless, we find that under certain circumstances (discussed in Secs.~\ref{sec::diffusion}~B and~C), a careful engineering of the environment can, in principle, improve the performance of both linear and second-order estimation schemes affected by phase diffusion. There remains an open question whether it is possible to obtain a similar improvement in the case of photon loss. This problem is now under investigation and is a subject for future work.

\begin{acknowledgements}\noindent
We gratefully acknowledge valuable discussions with Michael Hall, Marcin Jarzyna and Pieter Kok. The research leading to these results has received funding from the ARC Centre of Excellence CE110001027 and the European Union Seventh Framework Programme (FP7/2007-2013) under grant agreement n$^{\circ}$ [316244].
\end{acknowledgements}

\appendix*

\section{Full analytic form of $F^\text{min}_{Q}[|\Psi_{S,E}(\phi) \rangle]$}
Here we present the full analytic form of $F^\text{min}_{Q}[|\Psi_{S,E}(\phi) \rangle] = \min_{\{E_{k}(\phi)\}} F_{Q}[|\Psi_{S,E}(\phi) \rangle]$
\begin{widetext}
\begin{eqnarray}
F^\text{min}_{Q}[|\Psi_{S,E}(\phi) \rangle] &=& 4\eta \Bigl\{ \bigl[ \eta(\eta - 1)(2\eta - 1)^{2} \langle \hat{n} \rangle^{2}_{|\psi\rangle_{S}} + \bigl(6\eta(\eta - 1) + 1\bigr)(2\eta - 1)^{2} \langle \hat{n} \rangle_{|\psi\rangle_{S}} \nonumber \\
&-& \eta(\eta - 1)\bigl(7(2\eta - 1)^{2} \langle \hat{n}^{2} \rangle_{|\psi\rangle_{S}} - 4\eta(\eta - 1)\langle \hat{n}^{3} \rangle_{|\psi\rangle_{S}}\bigr) \bigr] \langle \hat{n}^{4} \rangle_{|\psi\rangle_{S}} \nonumber \\
&+& \eta(\eta - 1) \bigl[ \bigl(12\eta(\eta - 1) + 1\bigr) \langle \hat{n}^{3} \rangle_{|\psi\rangle_{S}} - 4\eta(\eta - 1)(\langle \hat{n} \rangle_{|\psi\rangle_{S}} + 6) \langle \hat{n} \rangle_{|\psi\rangle_{S}} - 4\langle \hat{n} \rangle_{|\psi\rangle_{S}} \bigr] \langle \hat{n}^{3} \rangle_{|\psi\rangle_{S}} \nonumber \\
&-& 2\eta(\eta - 1)\bigl[ 2\bigl(\eta(\eta - 1) + 1\bigr) + \bigl(8\eta(\eta - 1) + 1\bigr) \langle \hat{n} \rangle_{|\psi\rangle_{S}} \bigr] \langle \hat{n}^{3} \rangle_{|\psi\rangle_{S}} \langle \hat{n}^{2} \rangle_{|\psi\rangle_{S}} \nonumber \\
&-& \bigl[ 4\eta(\eta - 1)\bigl(\eta(\eta - 1)\bigl(\langle \hat{n}^{3} \rangle_{|\psi\rangle_{S}} - 4\bigr) - 1\bigr) + \bigl(2\eta(\eta - 1) + 1\bigr)(2\eta - 1)^{2}\langle \hat{n} \rangle_{|\psi\rangle_{S}} \bigr] \langle \hat{n}^{2} \rangle^{2}_{|\psi\rangle_{S}} \nonumber \\
&+& 7\eta(\eta - 1)(2\eta - 1)^{2} \langle \hat{n}^{2} \rangle^{3}_{|\psi\rangle_{S}} \Bigr\} \big{/} \Bigl\{ (1 - \eta)^{3} \left(\Delta \hat{n}^2\right)^{2}_{|\psi\rangle_{S}} - \eta(\eta - 1)\left[ 11\eta + 2(\eta - 1) \langle \hat{n} \rangle_{|\psi\rangle_{S}} - 4\right] \langle \hat{n}^{2} \rangle_{|\psi\rangle_{S}} \nonumber \\
&+& \eta\left[\langle \hat{n} \rangle_{|\psi\rangle_{S}} + (\eta - 1)\bigl(6(\eta - 1) \langle \hat{n}^{3} \rangle_{|\psi\rangle_{S}} + \eta(\langle \hat{n} \rangle_{|\psi\rangle_{S}} + 6)\langle \hat{n} \rangle_{|\psi\rangle_{S}} \bigr)\right] \Bigr\}\, .
\end{eqnarray}
\end{widetext}
Given this minimum and Eqs.~(\ref{CRbound}) and (\ref{minimum}) [with $m=1$] we can write an analytic lower bound on the error of the second-order estimation scheme affected by a one-arm photon loss valid for any input state
\begin{equation}\label{analyticbound}
\Delta \phi_{\rm LB} \geq \sqrt{\frac{1}{F_{Q}[\rho_{S}(\phi)]}} = \sqrt{\frac{1}{F^\text{min}_{Q}[|\Psi_{S,E}(\phi) \rangle]}}\, .
\end{equation}
For squeezed vacuum states containing the average number of photons $N$ [$\langle \hat{n} \rangle_\text{sv} = N$] $F^\text{min}_{Q}[|\Psi_{S,E}(\phi) \rangle]$ can be written as [the other required moments are given in Eqs.~(\ref{sv4}), (\ref{sv3}) and (\ref{sv2})]
\begin{widetext}
\begin{eqnarray}
F^\text{min}_{Q}[|\Psi_{S,E}(\phi) \rangle] &=& 32 N \eta \Bigl\{ N \bigl[ \eta(\eta - 1) \bigl(16\eta(\eta - 1) + 69\bigr) - 10\bigr] - 3 N^{2} \bigl[ 3 \eta(\eta - 1) \bigl(18\eta(\eta - 1) - 31\bigr) + 7\bigr] \nonumber \\
&-& 6 N^{3} \bigl[ 73 \eta \bigl( \eta \left[2 \eta(\eta - 2) + 1\right] + 1\bigr) + 2\bigr] - 6 N^{4} \eta(\eta - 1) \bigl[ 236 \eta(\eta - 1) - 37\bigr] \nonumber \\
&-& 720 N^{5} \eta^{2}(\eta - 1)^{2} + 6\eta(\eta - 1) - 1 \Bigr\} \nonumber \\
&\big{/}& \Bigl\{ 4 N (\eta - 1) \left[ 24 N^{2} (\eta - 1)^{2} + 21 N (\eta - 2)(\eta - 1) + \eta(2\eta - 17) + 20 \right] + 7 \eta - 8 \Bigr\}\, . \label{Fmin_sv}
\end{eqnarray}
\end{widetext}

\end{document}